\documentclass[ reprint, aps, prl, two column, showkeys]{revtex4-1}
\usepackage{graphicx,epsfig}
\usepackage{amssymb}
\usepackage{amsmath}
\usepackage{bm}
\usepackage{textcomp}
\usepackage{soul,xcolor}

\usepackage{color}

\begin{document}
\setcounter{page}{1}
\setstcolor{red}

\title[]{Record-quality GaAs two-dimensional hole systems}
\author{Yoon Jang \surname{Chung}}
\affiliation{Department of Electrical and Computer Engineering, Princeton University, Princeton, NJ 08544, USA  }
\author{C. Wang}
\affiliation{Department of Electrical and Computer Engineering, Princeton University, Princeton, NJ 08544, USA  }
\author{S. K. Singh}
\affiliation{Department of Electrical and Computer Engineering, Princeton University, Princeton, NJ 08544, USA  }
\author{A. Gupta}
\affiliation{Department of Electrical and Computer Engineering, Princeton University, Princeton, NJ 08544, USA  }
\author{K. W. \surname{Baldwin}}
\affiliation{Department of Electrical and Computer Engineering, Princeton University, Princeton, NJ 08544, USA  }
\author{K. W. \surname{West}}
\affiliation{Department of Electrical and Computer Engineering, Princeton University, Princeton, NJ 08544, USA  }
\author{R. \surname{Winkler}}
\affiliation{Department of Physics, Northern Illinois University, DeKalb, IL 60115, USA }
\author{M. \surname{Shayegan}}
\affiliation{Department of Electrical and Computer Engineering, Princeton University, Princeton, NJ 08544, USA  }
\author{L. N. \surname{Pfeiffer}}
\affiliation{Department of Electrical and Computer Engineering, Princeton University, Princeton, NJ 08544, USA  }

\date{\today}

\begin{abstract}

The complex band structure, large spin-orbit induced band splitting, and heavy effective mass of two-dimensional (2D) hole systems hosted in GaAs quantum wells render them rich platforms to study many-body physics and ballistic transport phenomena. Here we report ultra-high-quality (001) GaAs 2D hole systems, fabricated using molecular beam epitaxy and modulation doping, with mobility values as high as $5.8\times10^6$ cm$^2$/Vs at a hole density of $p=1.3\times10^{11}$ /cm$^2$, implying a mean-free path of $\simeq27$ $\mu$m. In the low-temperature magnetoresistance trace of this sample, we observe high-order fractional quantum Hall states up to the Landau level filling $\nu=12/25$ near $\nu=1/2$. Furthermore, we see a deep minimum develop at $\nu=1/5$ in the magnetoresistance of a sample with a much lower hole density of $p=4.0\times10^{10}$ /cm$^2$ where we measure a mobility of $3.6\times10^6$ cm$^2$/Vs. These improvements in sample quality were achieved by reduction of residual impurities both in the GaAs channel and the AlGaAs barrier material, as well as optimization in design of the sample structure.


\end{abstract}
\maketitle

\section{I. Introduction}

Charged carrier systems in a two-dimensional (2D) setting are useful for the study of ballistic transport phenomena as well as many-body physics. In sufficiently clean samples, the localization and scattering of carriers by impurities and defects are significantly suppressed, allowing them to traverse mesoscopic length scales without losing their initial information. Furthermore, the carrier density of such systems can be tuned so that the Coulomb energy dominates over the kinetic (Fermi) energy at very-low temperatures. Oftentimes, a perpendicular magnetic field is applied to the system to further enhance this tendency by discretizing the density of states into a series of Landau levels. A rich variety of interaction-driven phases have emerged in very-high-quality, modulation-doped GaAs 2D electron systems (2DESs) using this framework \cite{Tsui,Willett,Wigner1,Wigner2,Wigner3,stripe1,stripe2,ShayeganReview,JainBook,HalperinBook}.

The ground state of such correlated phases depends on the relative strength of the Coulomb interaction with respect to the other energies in the 2DES. Material parameters such as the effective mass, dielectric constant, and Land\'e $g$-factor play important roles here, since they determine the Fermi, cyclotron, Coulomb, and Zeeman energies. While GaAs 2DESs have been arguably the strongest leader for exploring many-body phenomena over the past few decades, quality permitting, there is substantial incentive to investigate other 2D systems with disparate material parameters. In this respect, GaAs 2D \textit{hole} systems (2DHSs) provide an exceptionally rich platform to study. When hosted in GaAs quantum wells (QWs), 2D holes have a much larger effective mass ($m^*_h$), even exceeding 1, compared to $m^*_e=0.067$ for GaAs electrons, both in units of free electron mass, $m_0$. This suggests that, at the same carrier density, interaction is comparatively stronger in GaAs 2DHSs compared to GaAs 2DESs as the larger mass lowers the Fermi energy. Moreover, the presence of heavy and light holes, anisotropic Fermi contours, as well as spin-orbit induced band splitting add further flavor to the physics of GaAs 2DHSs \cite{RolandBook}. 


These features have led to rigorous efforts to study the intricacies of interaction-driven phenomena in GaAs 2DHSs \cite{Mendez,MA1,MA2,MA10,15A,MB3,MA16,MA18,MA22,MA28,MA37,MB8,MB10,MA69,24A,MA72,MA74,MA78,MA81,MA85}. Examples include the observation and investigation of various magnetic-field-induced states such as the Wigner crystal and anisotropic stripe/nematic states \cite{MA1,MA2,15A,MA22,MB8,MB10,MA85}, single-layer and bilayer fractional quantum Hall state states \cite{Mendez,MA10,MA37,24A}, composite fermions with an anisotropic Fermi sea \cite{MA69,MA74,MA81}, and an unusual, even-denominator, fractional quantum Hall state \cite{MA72}, as well as an anisotropic Wigner crystal at a Landau level crossing \cite{MA78}. Thanks to their large effective mass and the ensuing large interaction parameter, $r_s$, defined as the ratio of the Coulomb to Fermi energy, GaAs 2DHSs have also been intensely studied at zero magnetic field in the context of the anomalous metal-insulator-transition problem in dilute 2D carrier systems \cite{MB3,MA18,MA28}, and a quantum Wigner crystal \cite{MA16}. Besides being systems of interest for many-body phases, GaAs 2DHSs' long mean-free-path and strong and tunable spin-orbit interaction have also rendered them prime candidates for studies of ballistic transport and spintronic phenomena \cite{MBB2,MAA3,MAA5,MAA8,MAA15,MAA21,MAA26,MAA30,MBB4,MAA38,MBB11,MAA80,MAA82,MBB12}. 

For all the above mentioned studies, it is crucial that sample quality is as high as possible so that subtle features can materialize in measurements without being hindered by disorder. Recently, there was a major breakthrough in the quality of GaAs 2DES that derived from a systematic reduction of background impurities via source purification and improvements in vacuum of the molecular beam epitaxy (MBE) growth chamber \cite{HighMobility}. It is then timely to evaluate the status of GaAs 2DHSs that are grown in the same MBE chamber and hosted in practically identical structures. Here we report the MBE growth of record-high-quality GaAs 2DHSs with mobility values as high as $\mu=5.8\times10^6$ cm$^2$/Vs at a temperature of $T=0.3$ K when the 2D hole density is $p=1.3\times10^{11}$ /cm$^2$. We also find an extremely high mobility of $\mu=3.6\times10^6$ cm$^2$/Vs in a sample with a much lower density of $p=4.0\times10^{10}$ /cm$^2$. Previously the highest mobilities achieved for GaAs 2DHSs at similar densities and temperature were $\mu=1.3\times10^6$ cm$^2$/Vs at $p\simeq1.3\times10^{11}$ /cm$^2$ and $\mu=2.3\times10^6$ cm$^2$/Vs at $p\simeq6.5\times10^{10}$ /cm$^2$ \cite{Manfra4}. We find that, in addition to the reduction of background impurities, optimizing the QW width is also essential in obtaining these results because it is closely correlated with the effective mass of the GaAs 2DHS.


\section{II. Experimental procedures}

Our 2DHSs are hosted in GaAs QWs fabricated on (001)-oriented GaAs substrates. The samples, whose typical structure is shown in Fig. 1(b), are grown by MBE, where our growth temperature is $T\sim650$ \textdegree C and we use C as the acceptor in an Al$_x$Ga$_{1-x}$As barrier to obtain holes in the GaAs QW via modulation doping \cite{ManfraHoles,WegsHoles}. The dopant is delivered to the growth space by heating a dog-bone-shaped filament of vitreous C that is heated through Ta leads with a power of $\sim200$ W, which generates a C acceptor delivery rate of $\sim10^{10}$ /cm$^2$s on the substrate. Typically our samples are $\delta$-doped for $\sim1$ to 3 minutes under these conditions, and the C filament is shuttered and kept at low power ($<1$ W) during the growth of undoped regions. When performing $\delta$-doping in the sample, growth is stopped and the temperature of the substrate is reduced to $T\lesssim500$ \textdegree C. Simultaneously, the power to the filament is raised to 200 W over a period of 100 s, and then the shutter is opened to introduce the C atoms into the AlGaAs barrier. After the doping is complete, the shutter is closed while the filament power is lowered again, and the growth of the AlGaAs barrier is resumed at $T\lesssim500$ \textdegree C. Following the growth of $\sim4.5$ nm of AlGaAs, the growth temperature is raised back to $ T\sim650$ \textdegree C. This scheme is implemented to minimize the surface segregation of C dopants during the growth of the structure. After growth, we cleave the wafers into $4\times4$ mm$^2$ pieces, and form In/Zn contacts to the 2DHS by thermal annealing at 450 \textdegree C for 4 minutes in a forming gas (N$_2$:H$_2$=95:5) environment. The samples are then cooled down to cryogenic temperatures for magnetotransport measurements in the dark.

\section{III. Results and discussion}

		
 \begin{figure*}[t]
\centering
    \includegraphics[width=.90\textwidth]{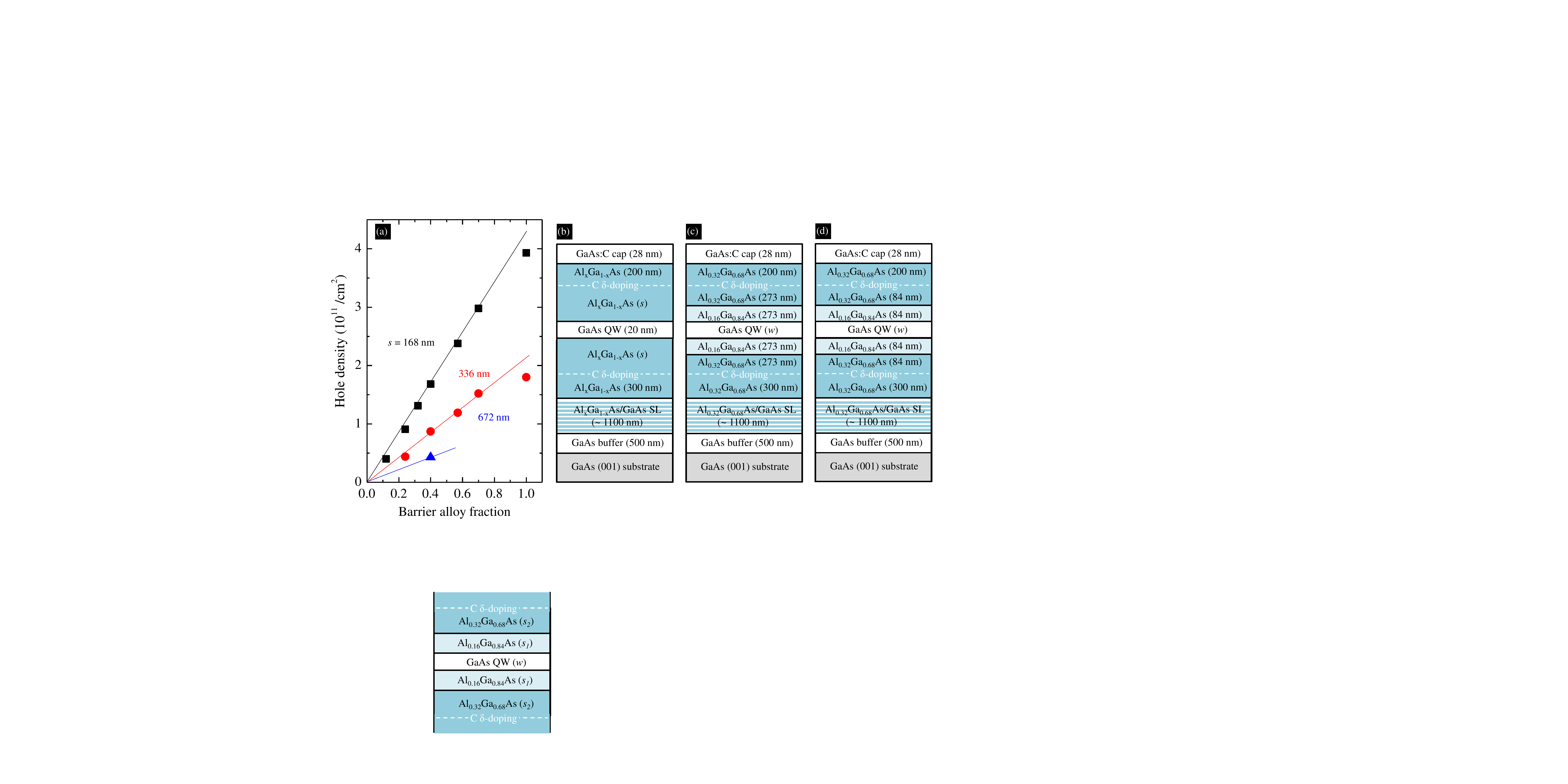}

 \caption{\label{fig1} (a) Two-dimensional hole density in GaAs QWs flanked by Al$_x$Ga$_{1-x}$As barriers with varying alloy fraction ($x$). Symbols with different colors show data from a series of samples grown with spacer layer thicknesses $s=168$ nm (black squares), 336 nm (red circles), and 672 nm (blue triangle). The solid lines are guides to the eye highlighting a linear dependence of hole density on $x$ for a given spacer layer thickness. (b) Schematic diagram of the structure used for the samples shown in (a). After a GaAs buffer layer, a superlattice (SL) of alternating Al$_x$Ga$_{1-x}$As (10 nm)/GaAs (1 nm) layers are grown on the substrate prior to defining the main region that hosts the 2DHS. All samples are $\delta$-doped with C symmetrically on both sides of the QW. The cap layer has a bulk C doping density of $\sim4\times10^{17}$ /cm$^3$. The structure of the high-mobility, stepped-barrier samples used for the 2DHSs discussed in Fig. 2 for (c) $p\simeq4.0\times10^{10}$ /cm$^2$ and (d) $p\simeq1.3\times10^{11}$ /cm$^2$. }
\end{figure*}	

\begin{figure*}[t]
 
 \centering
    \includegraphics[width=.98\textwidth]{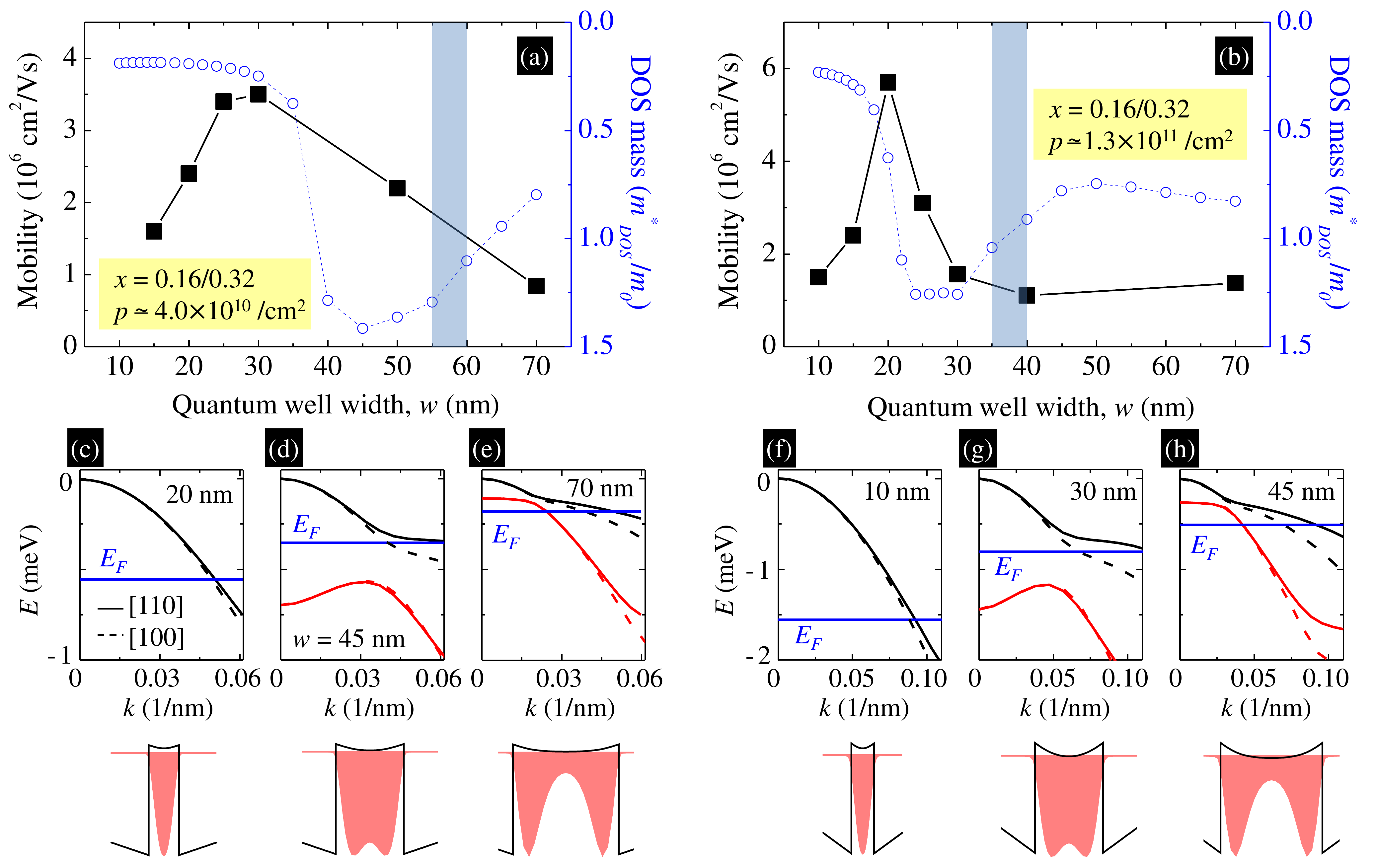} 
  \caption{\label{fig2} Mobility, measured at a temperature of 0.3 K, of the high-quality 2DHSs hosted in GaAs QWs with varying well width. The data in (a) are for a series of samples with density $p\simeq4.0\times10^{10}$ /cm$^2$ and in (b) for samples with $p\simeq1.3\times10^{11}$ /cm$^2$. The density variation is within $\pm5$\% for both sets, and the mobility for every data point is deduced from the measured density and resistance of each sample. The closed black squares denote experimentally measured mobility values and the open blue circles denote the calculated density-of-states effective masses for the ground-state subband. The light-blue band indicates the QW width range where, according to calculations, the population of the second subband is expected to begin. (c)-(e) and (f)-(h) show the calculated energy-band dispersions along the $[110]$ and $[100]$ directions at various QW widths for samples with $p=4.0\times10^{10}$ and $1.3\times10^{11}$ /cm$^2$, respectively. In each panel, the black and red curves represent the first and second subbands, while the Fermi energy ($E_F$) is marked by a solid blue line. The solid and dashed lines denote the dispersions along $[110]$ and $[100]$, respectively. The self-consistently calculated hole charge distribution functions (shaded red) and QW potential (black lines) for each of the cases of (c)-(h) are also shown at the bottom of each panel.}
\end{figure*} 

	
We first discuss the modulation-doping characteristics of C-doped GaAs 2DHSs. This information establishes a foundation for high-quality sample design. Figure 1(a) shows the GaAs 2DHS density as a function of Al$_x$Ga$_{1-x}$As barrier alloy fraction ($x$) for various spacer layer thicknesses. All samples are symmetrically doped from both sides of the QW, as shown in detail in Fig. 1(b), and the densities are measured by evaluating the quantum Hall features of magnetoresistance traces at $T=0.3$ K. Our modulation-doped GaAs 2DHSs show a steady increase in carrier density as the barrier Al alloy fraction is increased for all spacer layer thicknesses. 

The density of a modulation-doped 2DHS hosted in a QW can be estimated via the expression \cite{Davies}: 
\begin{align}
  \Delta E_v=E_0+E_F+E_A+E_{cap}.
   \label{eqn:1}
 \end{align}
Here $\Delta E_v$ is the valence-band offset at the QW/barrier interface, $E_0$ is the ground-state energy of the QW, $E_F$ is the Fermi energy with respect to $E_0$, $E_A$ is the acceptor level energy with respect to the valence-band edge, and $E_{cap}=pse^2/\epsilon_b\epsilon_0$ is the capacitive energy of the 2DHS; $p$, $s$, $e$, $\epsilon_b$, and $\epsilon_0$ are the 2DHS density, spacer layer thickness, fundamental electron charge, dielectric constant of the barrier, and vacuum permittivity, respectively.

Because of the relatively large effective mass of holes in GaAs, $E_0$ and $E_F$ are on the order of 1 meV in typical GaAs 2DHSs, making them negligibly small compared to $E_A$ and $E_{cap}$ in Eq. (1). Assuming that there is no drastic change in $E_A$ as the barrier $x$ varies, the 2DHS density is primarily determined by $\Delta E_v$. The monotonic increase in GaAs 2DHS density with the barrier $x$ is then reasonable since the band gap and hence $\Delta E_v$ is expected to gradually increase as the barrier becomes more Al rich. This is in contrast to modulation-doped GaAs 2DESs, where the carrier density shows a non-monotonic behavior as the barrier Al fraction is increased because of the $\Gamma$- to X-band crossover of the conduction band at $x\simeq0.4$ \cite{Adachi,DesignRules}.

\begin{figure*}[t]

\centering
    \includegraphics[width=.90\textwidth]{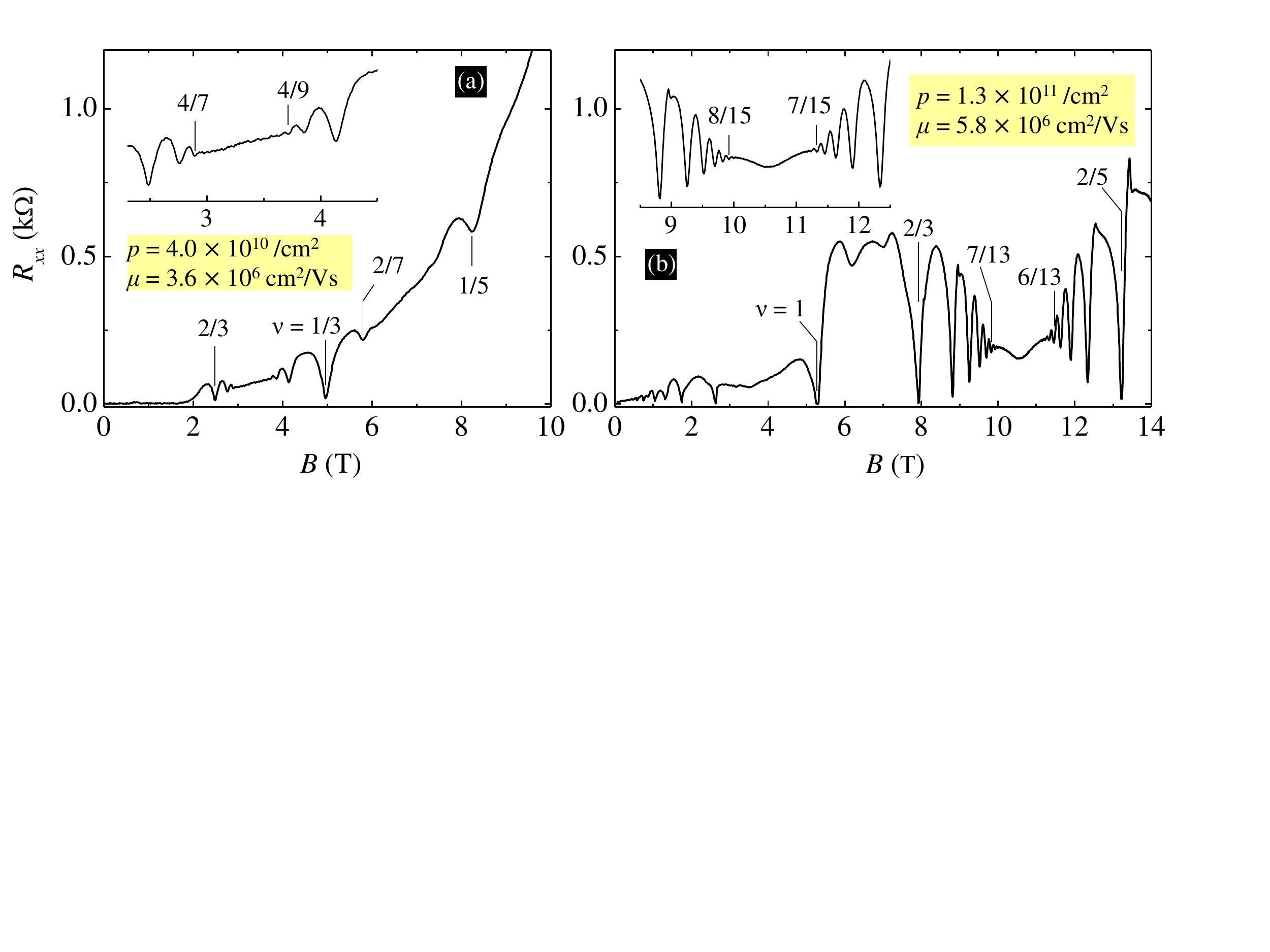} 
  \caption{\label{fig3} Representative magnetoresistance traces for high-quality 2DHSs with: (a) QW width $w=30$ nm and $p=4.0\times10^{10}$ /cm$^2$, and (b) $w=20$ nm and $p=1.3\times10^{11}$ /cm$^2$. All traces were taken at $T=0.3$ K in the van der Pauw geometry. The inset of each figure enlarges the region near $\nu=1/2$. The magnetic field positions of some integer and fractional quantum Hall states are marked.}
\end{figure*} 

It is possible to estimate a composition-dependent valence-band offset at the QW/barrier interface from the data plotted in Fig. 1(a). The density $p$ shows a fairly linear dependence on $x$: $p\simeq x\times4.4\times10^{11}$ /cm$^2$ when $s=168$ nm, and $p\simeq x\times2.2\times10^{11}$ /cm$^2$ when $s=336$ nm. This corresponds to a capacitive energy of $E_{cap}\simeq x\times517$ meV. Previously, $E_A$ levels have been reported to be $26\sim38$ meV for C in Al$_x$Ga$_{1-x}$As with $x\leq0.23$, showing a monotonic increase with $x$ \cite{CLevel}. Using these $E_{cap}$ and $E_A$ values along with Eq. (1), we deduce that $\Delta E_v$ changes with $x$ at a rate of $d\Delta E_v/dx\simeq560$ meV, which is reasonably consistent with what has been reported in the literature \cite{Adachi,Valence}.


We note that when the barrier is pure AlAs ($x=1.0$), the GaAs 2DHS density deviates from the linear trend observed in samples with lower-$x$ barriers. We are unsure of the reason for this behavior. Considering that we could not further increase the hole density even when significantly increasing the dopant density (data not shown), we speculate that there might be significant bowing of the valence band or C acceptor level in Al$_x$Ga$_{1-x}$As when $x>0.8$. Nevertheless, the data in Fig. 1(a) clearly demonstrate that very high density samples can be achieved in GaAs 2DHSs by performing modulation doping in high $x$ barriers.

\begin{figure*}[t]

\centering
    \includegraphics[width=.85\textwidth]{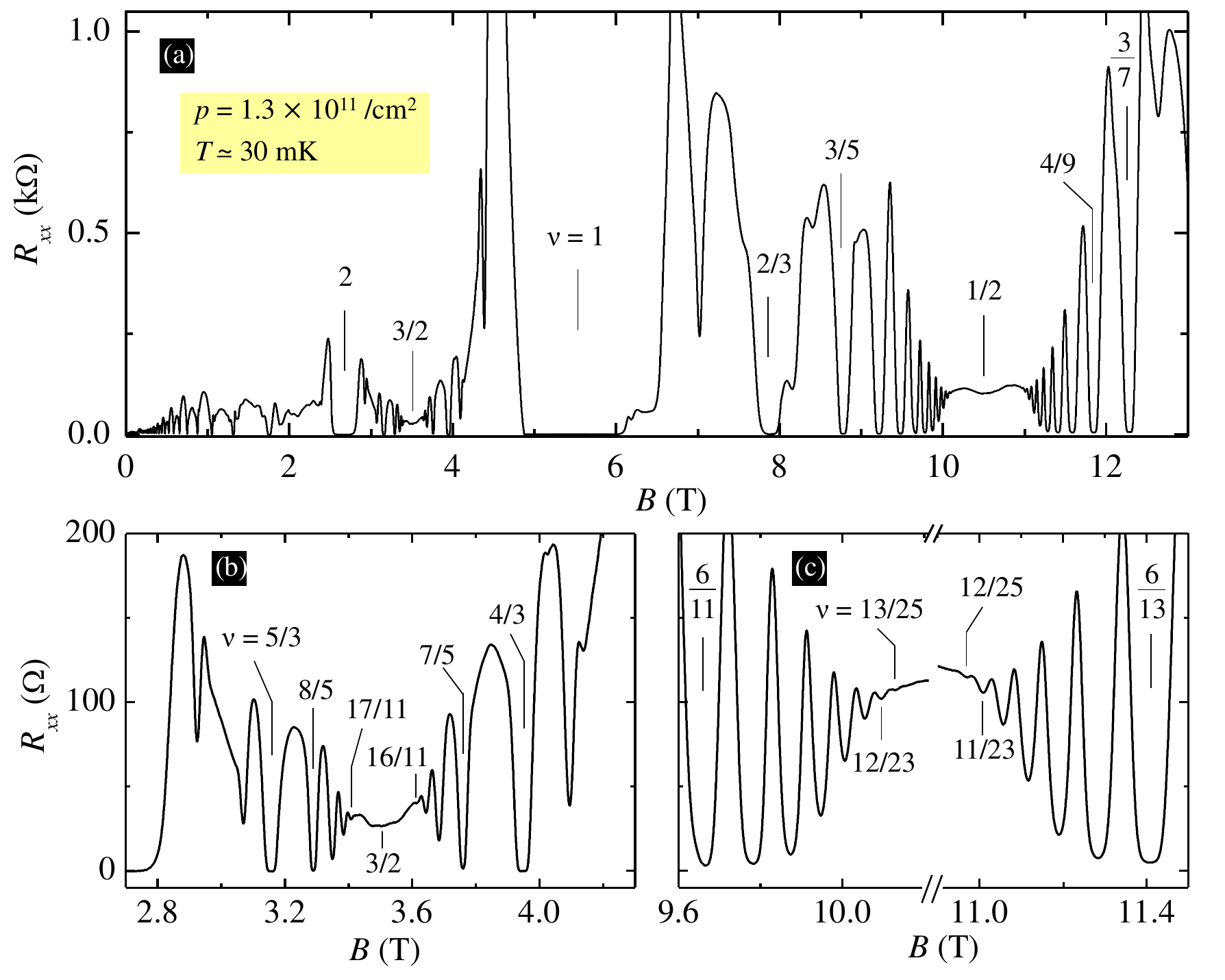} 
  \caption{\label{fig4} Representative magnetoresistance trace of the record-quality 2DHS with $p=1.3\times10^{11}$ /cm$^2$ and $w=30$ nm taken at $T\simeq30$ mK. (a) shows the data over the entire magnetic field range, while (b) and (c) enlarge regions near the Landau level fillings $\nu=3/2$ and 1/2, respectively. The magnetic field positions of several quantum Hall states, as well as $\nu=3/2$ and 1/2, are marked in each trace.}
\end{figure*} 
	
	While high carrier densities are generally beneficial for sample quality and/or mobility thanks to increased screening of charged impurities, other factors such as alloy, interface roughness, and second subband scattering should also be considered when aiming to design the best samples. The non-parabolic band structure that comes from mixing of the heavy- and light-hole bands makes the problem even more nuanced for GaAs 2DHSs \cite{RolandBook}. There have been a number of studies that have performed comprehensive analyses on optimizing the quality of such GaAs 2DHSs \cite{ManfraHoles,WegsHoles,Manfra4,Manfra2,Manfra3}. In the remainder of this paper, we provide an important update to these reports following the recent breakthrough in ultra-high-quality GaAs 2DESs made possible by a systematic impurity reduction in the MBE growth environment \cite{HighMobility}, which, as mentioned earlier, also implies an enhancement in GaAs 2DHS quality.
	
	Figures 2 (a) and (b) show the mobility vs. QW width ($w$) for a series of high-quality GaAs 2DHSs we prepared with two different densities. What first catches the eye are the extremely high peak mobilities of $\mu=3.6\times10^6$ and $\mu=5.8\times10^6$ cm$^2$/Vs for the samples with $p\simeq4.0\times10^{10}$ and $1.3\times10^{11}$ /cm$^2$, respectively. These represent remarkable improvements on the best previous mobility values reported for GaAs 2DHSs \cite{Manfra4}. This advancement comes from the growth of cleaner GaAs and AlGaAs material \cite{HighMobility}, as well as optimization of the sample design. Our samples implement a stepped barrier structure, where the alloy fraction of the barrier in the vicinity of the QW is $x=0.16$ while it is $x=0.32$ elsewhere (see Figs. 1(c) and (d)). This design was essential in realizing the high mobility values mentioned above. For example, the peak mobility degrades to  $\mu=2.9\times10^6$ cm$^2$/Vs  for the high density case and $\mu=1.6\times10^6$ cm$^2$/Vs for the low density case when a uniform barrier of $x=0.32$ is implemented instead of the $x=0.16/0.32$ stepped barrier. Furthermore, we find that it is also important to select an optimal QW width for each specific 2DHS density to achieve maximum mobility. This is because $w$ and $p$ both have a significant impact on the effective mass of the 2DHS hosted in the GaAs QW.



	An important feature of our samples is that the barriers are grown from extremely clean Al source material \cite{HighMobility,AlProblem}. Impurity concentrations are generally thought to be much higher in the Al$_x$Ga$_{1-x}$As barrier than in the GaAs QW for standard GaAs/Al$_x$Ga$_{1-x}$As heterostructures because of the high reactivity of Al with reagents such as O$_2$ or H$_2$O, and it is plausible that Al purity could have a meaningful impact on sample quality. Furthermore, in finite-barrier-potential samples with narrow QW width, which are typical for high-mobility GaAs 2DHSs, there is significant wave function penetration into the barrier, which could aggravate the influence of high impurity concentration in the barrier. We measure mobility values comparable to the best previous GaAs 2DHS that use barrier alloy compositions $x\leq0.16$ even in our samples that use a uniform barrier of $x=0.32$ for the entire structure \cite{Manfra4}. This attests to the extraordinary cleanliness of the Al source in our MBE chamber.


Next, we comment on the non-monotonic evolution of mobility vs. $w$ in Figs. 2(a) and (b). The qualitative trend of mobility peaking at a relatively narrow well width and then rapidly dropping as $w$ increases, seen in both series of samples, is qualitatively similar to what has been observed previously \cite{ManfraHoles,Zhu}. For narrow well widths ($w\leq20$ nm) the mobility increases with $w$ most likely because of a decrease in interface roughness scattering but, as $w$ increases further, the complex evolution of the effective mass also becomes important. The blue open circles in Figs. 2(a) and (b) denote the density-of-states (DOS) effective mass ($m^*_{DOS}$) values at the Fermi energy deduced from the hole energy-band dispersions we calculate for the ground-state subband as a function of $w$. Here, our self-consistent calculations are based on an $8\times8$ Kane Hamiltonian with cubic anisotropy but no Dresselhaus term \cite{RolandBook}. Figures 2(c)-(e) show the calculated band dispersions for some representative QW widths when $p=4.0\times10^{10}$ /cm$^2$ and Figs. 2(f)-(h) show similar examples for when $p=1.3\times10^{11}$ /cm$^2$. Here the solid and dashed lines indicate the dispersion relations in the $[110]$ and $[100]$ directions, while the colors black and red are used to specify the first and second subbands, respectively. The blue solid line represents the Fermi energy in each panel.

If we compare the GaAs 2DHS mobility and $m^*_{DOS}$ values plotted in Figs. 2(a) and (b), it is evident that, regardless of the 2DHS density, the mobility starts to exhibit a noticeable decline at well widths where $m^*_{DOS}$ begins to increase. Qualitatively, this is reasonable considering that we expect the mobility to decrease when the effective mass increases in 2D carrier systems \cite{Davies}. Note that according to our calculations, the onset of the second subband, denoted by the light-blue bands in Figs. 2(a) and (b), occurs well after the mobility has decreased significantly. This rules out the occupation of the second subband as the primary culprit for the rapid drop in mobility observed in the data as $w$ is increased. In the calculation, the broad plateau of $m^*_{DOS}$ as a function of well width is due to the fact that the Fermi energy is pushed through an anticrossing between the highly nonparabolic dispersion of the first and second subband.  We speculate that scattering contributions from the second subband are indeed responsible for the absence of a recovery in the mobility trend at larger $w$ values where $m^*_{DOS}$ starts to decrease again. At even wider well widths where the second subband is significantly occupied, we suspect that the mobility should increase again since our calculations indicate that in general $m^*_{DOS}$ for the second subband is small compared to that of the ground-state subband. One could argue that a glimpse of this behavior is starting to show for the 70 nm sample in Fig. 2(b), although additional complications from the bilayer-like charge distribution could also be causing aberrations in this case. All in all, the data shown in Fig. 2 demonstrate that optimizing the QW width is crucial in obtaining the highest-quality GaAs 2DHS with a specific carrier density.

It is worth comparing the peak-mobility values of our GaAs 2DHSs with those of state-of-the-art GaAs 2DESs \cite{HighMobility}. Assuming that scattering contributions from residual background impurities determine the peak mobility in our GaAs 2DHSs and that material cleanliness is similar for the GaAs 2DESs and 2DHSs grown in our chamber, we can estimate the expected GaAs 2DHS mobility. Using simple models \cite{Davies}, we obtain a residual background impurity concentration of $1\times10^{13}$ /cm$^3$ for GaAs 2DESs \cite{HighMobility}. Applying the same model to GaAs 2DHSs, we find that the expected mobilities are still a factor of $\sim5$ larger than what we measured for our peak-mobility samples for the high-density samples in Fig. 2 at $T=0.3$ K. For the low-density samples, the mobility is off by a slightly smaller factor of $\sim4$. 

We point out that such simple models work only in the zero-temperature limit. Unlike GaAs 2DESs, screening of the ionized impurities by the holes and hence the mobility in GaAs 2DHSs, can be temperature dependent even when $T\leq1$ K \cite{MB3,Manfra3,DasSarmaScreening}. It is plausible then that the peak mobility of the GaAs 2DHSs measured at $T=0.3$ K is lower than what is expected for zero temperature since screening would be weaker at finite temperatures. Indeed, it has been shown that the mobility of GaAs 2DHSs with $p\leq8\times10^{10}$ /cm$^2$ decreases by a factor of $\simeq1.5$ when the temperature was raised from $T\simeq50$ mK to 0.3 K \cite{MB3,Manfra3}.
However, we would like to note that the mobility of our high-density sample varies less than 5\% when comparing measurements at $T\simeq30$ mK and 0.3 K. This is not inconsistent with previous results since the density is much higher in our case and the temperature dependent change in resistance for GaAs 2DHSs has been shown to decrease as the carrier concentration increases \cite{MB3}. Another potential explanation for the mobility disparity mentioned above is that the charge conditions of impurities could depend on the position of the Fermi level. In this scenario, even after considering the difference in effective mass, the scattering caused by residual background impurities would be different for GaAs 2D electron and hole systems regardless of them residing in samples with similar structure and purity.

Of course, we cannot rule out the possibility that other scattering mechanisms effect the GaAs 2DHS mobility in our samples. While the larger effective masses of GaAs 2DHSs imply a weaker influence of interface roughness scattering compared to GaAs 2DESs at zero temperature \cite{LiIR}, finite-temperature corrections to screening could alter these results and make interface roughness scattering relevant in our case. This is supported by the fact that in Fig. 2, we see a substantial increase in mobility as $w$ increases when the QW is narrow, which mimics what has been observed for GaAs 2DESs where interface roughness scattering determines the mobility \cite{Sakaki,DobyMob}. One should also keep in mind that the spin-orbit interaction and inversion asymmetry in GaAs/AlGaAs heterostructures lead to spin-split subbands \cite{RolandBook}. Such a splitting is particularly significant for GaAs 2DHSs \cite{RolandBook,MBB2,MAA15,MAA30,MBB4,MAA38,MBB11}. Given this circumstance, it is conceivable that spin-subband scattering is playing a role although, as has been pointed out previously \cite{Manfra4}, it is not straightforward to imagine what could provide a spin-flip mechanism at the low temperatures we perform our measurements. Finally, we cannot fully exclude the possibility that there are more impurities in our GaAs 2D hole samples compared to our electron samples. In terms of MBE growth, the primary difference between GaAs 2DHSs and 2DESs is the type of dopant used to obtain carriers in the channel. We use C for holes and Si for electrons. Surface segregation is well understood for Si in MBE-grown GaAs/AlGaAs \cite{Si1,Si2,Si3,Si4}, but is still rather obscure for C. It is possible that dopant migration is much more severe in our hole samples despite the preventive measures we take by lowering the temperature of the substrate when introducing the C dopants. More subtle differences such as the larger amount of electric power required to generate a sufficient C flux compared to Si during doping could also cause relatively higher impurity concentrations in our GaAs 2D hole samples. Overall, despite the record-mobility results, at this time we are unsure as to what is limiting us from achieving even better mobility GaAs 2DHSs.


Given the major progress we have made in GaAs 2DHS mobility, it is also helpful to examine the magnetotransport characteristics of our samples. Figures 3(a) and (b) show magnetoresistance ($R_{xx}$) traces of the highest mobility samples for GaAs 2DHSs with $p=4.0\times10^{10}$ and $p=1.3\times10^{11}$ /cm$^2$, respectively. Clearly, the quality of our samples is extraordinary. In the $R_{xx}$ trace of the low-density sample shown in Fig. 3(a), there are clear signs of four-flux fractional quantum Hall states (FQHSs) developing at $\nu=1/5$ and 2/7 even though the measurement is only performed at the moderately low temperature $T=0.3$ K. This is impressive considering that a much lower temperature was required to observe signatures of these fragile states even in the best previous GaAs 2DHSs \cite{MA85}. The $R_{xx}$ trace of the high-density sample shown in Fig. 3(b) is equally remarkable, as high-order FQHSs are observed up to Landau level fillings $\nu=8/15$ and 7/15 on the flanks of $\nu=1/2$.

The extraordinary quality of our GaAs 2DHSs becomes even more evident when the samples are measured at dilution refrigerator temperatures. Figure 4 shows the $R_{xx}$ trace of our $p=1.3\times10^{11}$ /cm$^2$ sample measured at $T\simeq30$ mK. It is already clear from the full-field trace shown in Fig. 4(a) that the sample exhibits a multitude of FQHSs in $R_{xx}$, especially near $\nu=3/2$ and 1/2. Figures 4(b) and (c) enlarge these regions, and in this expanded view it is possible to discern very-high-order FQHSs such as $\nu=17/11$ and 16/11 near $\nu=3/2$, and $\nu=13/25$ and 12/25 near $\nu=1/2$, developing in $R_{xx}$. These are the highest-order FQHSs ever reported for any 2DHS \cite{15A,Manfra4,ManfraHoles,Manfra3}. Nevertheless, they are not as high-order as the highest seen in the best quality 2DESs \cite{HighMobility}. It is possible that this is partly because of the potentially larger amount of disorder in 2DHSs, as already discussed above. However, it is more likely that the culprit is the larger effective mass of holes as it results in a smaller separation between the Landau levels. This in turn leads to a more severe Landau level mixing, which is known to weaken and lower the energy gaps of FQHSs \cite{JainBook,MA1,MA2,MA85,Kevin}.


\section{IV. Summary and outlook}

In summary, we report the growth of very-high-quality (001)-oriented GaAs 2D hole samples. The modulation-doping characteristics and design parameters of our samples are described in detail, and we establish a foundation for high-quality GaAs 2DHS preparation. In an optimized structure, we measure a record-high mobility value of $\mu=5.8\times10^6$ cm$^2$/Vs in a GaAs 2DHS with a density of $p=1.3\times10^{11}$ /cm$^2$. We conjecture that minimizing residual impurities not only in the channel but also in the barrier was crucial in achieving these results. Several delicate FQHSs are observed in the magnetoresistance traces of our representative samples at $T=0.3$ K, and our record-mobility sample displays unprecedented quality when measured at $T\simeq30$ mK with high-order FQHSs such as $\nu=12/25$ and 13/25 emerging near $\nu=1/2$. Such signatures strongly suggest that the improvement we present here will prove useful for future studies of many-body physics in GaAs 2DHSs.

\begin{acknowledgments}
We acknowledge support by the National Science Foundation (NSF) Grant Nos. DMR 2104771, ECCS 1906253, and MRSEC DMR 2011750, the Eric and Wendy Schmidt Transformative Technology Fund, and the Gordon and Betty Moore Foundation EPiQS Initiative (Grant No. GBMF9615 to L.N.P.) for sample fabrication and characterization. For measurements, we acknowledge support by the U.S. Department of Energy Basic Energy Sciences (Grant No. DEFG02-00-ER45841).
 \end{acknowledgments}
 
 The data that support the findings of this study are available from the corresponding author upon request.

\end{document}